\documentclass[prb,aps,11pt]{revtex4}
\usepackage{epsf}
\usepackage{times}
\def\rr{{\bf r}}

\def\k{{\bf k}}
\def\b{{\bf b}}
\def\a{{\bf a}}
\def\G{{\bf G}}
\newcommand{\mbeg}{\begin{eqnarray}}
\newcommand{\mend}{\end{eqnarray}}
\newcommand{\R}{{\bf R}}
\newcommand{\F}{{\bf F}}
\newcommand{\idr}{\int\! d{\bf r}\,}

\newcommand{\bk}[2] {\langle #1 | #2 \rangle }
\newcommand{\pard}[2]{\frac{\partial #1 }{ \partial #2 } }
\newcommand{\var}[2]{\frac{\delta #1 }{ \delta #2 } }

\def\be{\begin{equation}}
\def\ee{\end{equation}}
\def\beq{\begin{equation}}
\def\eeq{\end{equation}}

\def\ket#1{| #1 \rangle}
\def\bra#1{\langle #1 |}

\def\bnm{_{nm}}

\def\bloc{_{{\rm loc}}}

\def\bNL{_{\rm NL}}
\def\bh{_{\rm H}}
\def\bxc{_{\rm xc}}

\def\btot{_{\rm tot}}
\def\beff{_{\rm eff}}

\def\bks{_{\rm ks}}
\def\bsoft{_{\rm soft}}

\def\pion{^{\rm \,ion}}

\def\po{^{(0)}}

\def\bcpn{_{\rm c}^{\,\rm dens}}
\def\bcpw{_{\rm c}^{\,\rm wf}}
\def\bcps{_{\rm c}^{\,\rm soft}}

\begin{document}

\title{First-principle molecular dynamics with 
ultrasoft pseudopotentials: parallel implementation
and application to extended bio-inorganic systems}

\author{P. Giannozzi$^{1,2}$, F. De Angelis$^{1,3}$, R. Car$^1$ \\
$^1$ Department of Chemistry and Princeton Materials Institute\\
    Princeton University, Princeton NJ 08544, USA\\
$^2$ NEST-INFM, Scuola Normale Superiore di Pisa, Italy  \\
$^3$ Istituto CNR di Scienze e Tecnologie Molecolari (ISTM), c/o\\
    Dipartimento di Chimica, Universit\`a di Perugia, Italy\\
}
\begin{abstract}

We present a plane-wave ultrasoft pseudopotential implementation
of first-principle molecular dynamics, which is well suited to model
large molecular systems containing transition metal centers.
We describe an efficient strategy for parallelization
that includes special features to deal with the augmented charge
in the contest of Vanderbilt's ultrasoft pseudopotentials.
We also discuss a simple approach to model molecular systems with
a net charge and/or large dipole/quadrupole moments.
We present test applications to manganese and iron porphyrins
representative of a large class of biologically relevant metallorganic
systems. Our results show that accurate Density-Functional Theory 
calculations on systems with several hundred atoms are feasible
with access to moderate computational resources.
\end{abstract}

\maketitle
\today

\newpage

\section{Introduction}

There is increasing interest in studying the electronic structure 
of complex biological molecules. This is an essential step to 
understand e.g. enzymatic and/or biomimetic catalysis. Modeling
bio-catalytic systems is however very challenging, because a proper
description of the active site needs the inclusion of a large number 
of atoms (from several tens to a few hundreds) treated at a
high level of quantum chemical theory. In this respect a good
compromise in terms of accuracy and computational cost is
provided by Density-Functional Theory (DFT)\cite{DFT}, whose use to
model the electronic structure of protein active sites is becoming 
increasingly popular.
In combination with Car-Parrinello (CP)\cite{CP} first-principle
molecular dynamics (MD), DFT allows us to optimize molecular structures,
study dynamical and finite-temperature properties, and model 
reaction paths.

In most standard implementations, the CP method
employs a plane-wave (PW) basis set. An advantage of PWs is
that they do not depend on atomic positions and are
free of basis-set superposition errors.
Total energies and forces on the atoms can be calculated
using computationally
efficient Fast Fourier transform (FFT) techniques.
Finally, the convergence of a calculation can be controlled in a
simple way, since it depends only upon the number of PWs
included in the expansion of the electron density.
The dimension of a PW basis set is controlled by
a cutoff in the kinetic energy of the PWs, which is usually 
measured in Ry units.
A disadvantage of PWs is
their extremely slow convergence
in describing core states. To deal with this
difficulty, one usually employs norm-conserving (NC)
pseudopotentials (PPs) \cite{NCPP}
to model the interaction of the valence electrons with the ionic
core (nucleus + core electrons). 
Parallel implementations of PW calculations
based on NC PPs are well documented in the 
literature (see e.g. \cite{ParaFFT,MarxHutter,Cavazzoni}). 

When using NC PPs, very large PW basis sets are
needed to accurately represent
the contracted $p$ orbitals of the first-row elements O, N, F,
and the $3d$ orbitals of the transition metal block.
These orbitals belong to elements for which the NC PPs
are ``hard'', typically requiring cutoffs of more than
70 Ry in order to yield sufficiently converged results. 
As a comparison, calculations on elements like Al, Si, P,
for which the corresponding NC PPs are ``soft'', 
are usually well converged with a cutoff of 20 Ry or less.
A consequence of the delocalized nature of the PWs is that
the presence of a single hard PP in a system
requires the use of a correspondingly high cutoff
for all the other PPs.
This difficulty is particularly serious for metallorganic
systems containing one or more transition metal centers.
The high cutoff required for such atoms translates into a very
large number of PWs, which in turn implies long execution times
and large memory requirements.

An approach that drastically reduces the PW cutoff
was proposed by Vanderbilt \cite{Van}, who introduced \
``ultrasoft'' (US) PPs. 
In this approach the normalized charge density is the sum of two
terms, a soft part represented in terms of smooth orbitals, and
a hard part which is treated as an augmented charge.
A closely related approach, the ``Projector-augmented wave''
(PAW) method introduced by Bl\"ochl \cite{Blo}, 
is an all-electron rather than a PP electronic structure method.
The PAW approach provides a simple and effective 
algorithm for reconstructing all-electron orbitals from 
pseudo-orbitals\cite{Reconstruct}.
An efficient serial implementation of the CP scheme
with US PPs is described in Ref. \cite{CarPasquarello}.

In this paper we present in detail a parallel 
implementation of the CP scheme using US PPs.
We also provide further details on the 
procedures used in serial implementation.
We compare the relative efficiency
of US and NC PPs in realistic calculations for large
molecules, performed on parallel machines.
Our test molecules are a reduced and an extended model 
of the active site of myoglobin, containing the iron-porphyrin motif.
We focus on metalloporphyrin systems because they are representative 
of a large class of 
biomolecules which can be modeled efficiently with US PPs.
Our benchmarks indicate that US calculations
are at least 2-3 times less expensive than NC calculations
of comparable accuracy.

The use of a PW basis set implies that Periodic Boundary Conditions
(PBC) are imposed, i.e. an isolated molecule has to be placed into a 
periodically repeated box (a ``supercell''). The supercell must be
large enough to ensure that the total potential is vanishingly small 
at the box boundary, thus minimizing spurious interactions between
periodic replicas. For neutral systems with small dipole/quadrupole
moments, supercells of reasonable size can be safely used.
For charged molecules, or molecules with large dipole/quadrupole 
moments, however, the error induced by PBC may be rather large
unless exceedingly large supercells are used.

We follow here a technique introduced by Makov and Payne 
(MP)\cite{MakovPayne} to eliminate the spurious electrostatic
interactions in the latter case. The MP technique is approximate
because it is not self-consistent and takes into account only
moments up to quadrupole. To check the accuracy of the MP
technique, we compare CP calculations on highly charged manganese 
porphyrins, performed using PWs and US PPs, with calculations on the same
systems using localized basis sets which do not require PBC.
The comparison shows that the MP correction yields results that
to all practical effects are indistinguishable from results
obtained without PBC.

The paper is organized as follows. In Sec.2, we recall the main
aspects of US-PP implementation in the serial case.
In Sec.3, we describe our parallel implementation.
In Sec.4, we compare the computer performances of US and NC
PPs for a 
reduced and an extended model of the myoglobin active site. 
In Sec.5, we compare CP calculations
with localized basis-set calculations not requiring PBC. The
test systems are highly-charged isomeric 
{\em meso}-substituted manganese porphyrins. 
Sec.6 contains our conclusions.

\section{Plane-Wave Ultrasoft Pseudopotential Implementation}

\subsection{Kohn-Sham equations with Ultrasoft Pseudopotentials}

The implementation of CP molecular dynamics within
a US-PP framework is described in Ref.\cite{CarPasquarello}.
Here we briefly remind the main formulas, using the same notation
of Ref.\cite{CarPasquarello}.

The total energy of a system of $N_v$ valence electrons, having
one-electron Kohn-Sham (KS) orbitals $\phi_i$, is given by
\mbeg
  E\btot [\{ \phi_i\},\{ \R_I\}] & = &
  \sum_i \bra{\phi_i} -\frac{\hbar^2}{2m}\nabla^2 + V\bNL \ket{\phi_i}
 + E\bh[n] + E\bxc[n] \nonumber\\
 & & \quad + \int\! d\rr\, V\bloc\pion(\rr) n(\rr) + U(\{\R_I \})
  \;\;,
\label{etot}
\mend
where $n(\rr)$ is the electron density, $ E\bh[n] $ is the Hartree 
energy:
\mbeg
 E\bh[n] = \frac{e^2}{2} \int\!\!\int\! d\rr\, d\rr'\, \frac{n(\rr) n(\rr')}
{|\rr -\rr'|}\;\;,
\mend
$E\bxc[n]$ is the exchange and
correlation energy, $U(\{\R_I\})$ is the ion-ion interaction energy,
and $\R_I$ indicate atomic positions.
In the following, potentials have energy dimensions.
The PP is composed of a local part $V\bloc\pion$, given by a sum
of atom-centered radial potentials:
\mbeg
V\bloc\pion(\rr) = \sum_I V\bloc^I(\,|\rr-\R_I|\,)
\mend
and a nonlocal part $V\bNL$:
\vskip 0.3cm
\mbeg
   V\bNL = \sum_{nm,I} D\po_{nm} \ket{\beta_n^I} \bra{\beta_m^I} \;,
\label{nonloc}
\mend
where the functions $\beta^I_n$ and the coefficients
$D\po_{nm}$ characterize the PP and are specific for 
each atomic species. For simplicity, we consider a single atomic 
species only in what follows. The $\beta_n^I$ functions,
centered at site $\R_I$, depend on the ionic positions through
\mbeg
\beta^I_n(\rr)=\beta_n(\rr-\R_I) \;.
\label{beta}
\mend
$\beta_n$ is an angular momentum eigenfunction in the angular
variables, times a radial function which vanishes outside the core
region; the indices $n$ and $m$ in Eq.\ (\ref{nonloc}) run over the
total number $N_\beta$ of these functions.

The electron density in Eq.\ (\ref{etot}) is given by
\mbeg
n(\rr)=\sum_i \Big[\; |\phi_i(\rr)|^2+\sum_{nm,I} Q_{nm}^I(\rr)
\langle\phi_i|\beta_n^I\rangle\langle\beta_m^I|\phi_i\rangle \; \Big]\;,
\label{dens}
\mend
where the sum runs over occupied KS orbitals.
The augmentation functions $Q^I_{nm}(\rr)=Q_{nm}(\rr-\R_I)$ are 
localized in the core.
The ultrasoft PP is fully determined by the quantities
$V\bloc^I(r), D_{nm}\po, Q_{nm}(\rr),$ and
$\beta_n(\rr)$.
The functions $Q_{nm}(\rr)$ are defined in terms of atomic orbitals
as: $Q\bnm(\rr) = \psi_n^{ae*}(\rr)\psi^{ae}_m(\rr)
                - \psi_n^{ps*}(\rr)\psi^{ps}_m(\rr)$
where $\psi^{ae}$ are atomic one-electron orbitals (not necessarily bound), 
and $\psi^{ps}$ are the corresponding pseudo-orbitals.
The $Q_{nm}(\rr)$ are pseudized as described in Ref.\cite{CarPasquarello}.
This enables us to treat the $Q_{nm}(\rr)$ with Fourier transform techniques.

The KS orbitals obey generalized orthonormality conditions
\mbeg
\bra{\phi_i}\, S(\{\R_I\}) \, \ket{\phi_j} = \delta_{ij} \;,
\label{ortho}
\mend
where $S$ is a Hermitian overlap operator given by
\mbeg
S=1+\sum_{nm,I} q_{nm} |\beta_n^I\rangle\langle\beta_m^I| \;,
\label{sop}
\mend
and
\mbeg
q_{nm}=\int d\rr \,Q_{nm}(\rr).
\mend
The orthonormality
condition (\ref{ortho}) is consistent with the conservation
of the charge $\int d\rr\, n(\rr)=N_v$. Note that the overlap
operator $S$ depends on ionic positions through the
$|\beta_n^I\rangle$.

The ground-state orbitals $\phi_i$ minimize
the total energy (\ref{etot}) subject to the constraints
(\ref{ortho}),
\mbeg
\var{E\btot}{\phi_i^*(\rr)} = \epsilon_i\, S \phi_i(\rr) \;,
\label{single}
\mend
where the $\epsilon_i$ are Lagrange multipliers.
Eq.(\ref{ks}) is the KS equation:
\mbeg
H \ket{\phi_i} = \epsilon_i\, S \ket{\phi_i} 
\label{ks}
\mend
where
\mbeg
H = -\frac{\hbar^2}{2m}\nabla^2 + V\beff + 
\sum_{nm,I} D_{nm}^I |\beta_n^I\rangle\langle\beta_m^I| \;.
\label{ham}
\mend
$V\beff$ is a screened effective local potential,
\mbeg
V\beff(\rr) = 
V\bloc\pion({\rr}) + V\bh(\rr) + \mu\bxc(\rr) \;.
\label{veff}
\mend
$\mu\bxc(\rr)$ is the exchange-correlation potential:
\mbeg
\mu\bxc(\rr) = {\delta E\bxc[n] \over \delta n(\rr)} \;,
\label{vxc}
\mend
and $V\bh(\rr)$ is the Hartree potential:
\mbeg
V\bh(\rr) = e^2 \int\! d\rr'\, \frac{n(\rr')}{ |\rr-\rr'|} \;.
\label{vh}
\mend
The ``screened'' coefficients $D_{nm}^I$ appearing in Eq.\ (\ref{ham})
are defined as:
\mbeg
D_{nm}^I = D_{nm}\po + \idr V\beff(\rr) Q^I_{nm}(\rr) \;.
\label{dij}
\mend
They depend on the KS orbitals through $V\beff$, Eq.\ (\ref{veff}), 
and the charge density, Eq.\ (\ref{dens}).

\subsection{Molecular dynamics with Ultrasoft Pseudopotentials}
\label{cpvan}

In the CP approach,\cite{CP} the electronic orbitals and
the ionic coordinates evolve according to a classical Lagrangian
\mbeg
{\cal L} = \mu \sum_i \int\! d\rr\, |\dot \phi_i(\rr)|^2 +
\frac{1}{2}\sum_I M_I \dot \R_I^2 - E\btot(\{\phi_i\},\{\R_I\}) \;,
\label{lagrange}
\mend
subject to a set of constraints
\mbeg
{\cal N}_{ij}(\{\phi_i\},\{\R_I\}) = \bra{\phi_i} S \ket{\phi_j}
     - \delta_{ij} = 0 \;.
\label{constraints}
\mend
Here $\mu$ is a fictitious mass parameter for the electronic degrees of
freedom, $M_I$ are the masses of the atoms, and $E\btot$ and $S$ are 
given in Eqs.\ (\ref{etot}) and (\ref{sop}), respectively.
The holonomic orthonormality constraints (\ref{constraints}) 
do not cause energy dissipation in an MD run.  They may be 
incorporated in the Euler equations of motion by introducing 
Lagrange multipliers $\Lambda_{ij}$:
\mbeg
  \mu \ddot \phi_i &=& - \var{E\btot}{\phi_i^*} +
\sum_j \Lambda_{ij}  S \phi_j \;,
\label{ele} \\
 \F_I=M_I \ddot \R_I &=& - \pard{E\btot}{\R_I} +
\sum_{ij} \Lambda_{ij} \bra{\phi_i} \pard{ S}{\R_I}
\ket{\phi_j} \;.
\label{ions}
\mend
At equilibrium, Eq.\ (\ref{ele}) reduces 
to the electronic KS equations (\ref{single}) or (\ref{ks}).
A unitary rotation brings the $\Lambda$ matrix into
diagonal form: $\Lambda_{ij}=\epsilon_i\delta_{ij}$.
The equilibrium for the ions is achieved when the ionic forces $\F_I$ 
in Eq.\ (\ref{ions}) vanish.

In deriving explicit expressions for the forces, Eq.\ (\ref{ions}), 
one should keep in mind that the electron density also depends on
$\R_I$ through $Q^I_{nm}$ and $\beta^I_n$. Introducing the
quantities
\mbeg
\rho_{nm}^I = \sum_i \bk{\phi_i}{\beta^I_n} \bk{\beta^I_m}{\phi_i} \;,
\mend
and
\mbeg
\omega_{nm}^I = \sum_{ij} \Lambda_{ij}
                     \bk{\phi_j}{\beta^I_n} \bk{\beta^I_m}{\phi_i} \;,
\mend
we arrive at the expression
\mbeg
 \F_I &=&
 - \frac{\partial U}{\partial \R_I}
 - \int d\rr\, \frac{\partial V\bloc\pion}{\partial\R_I} n(\rr)
 - \int d\rr\, V\beff(\rr) \sum_{nm}
               \frac{\partial Q_{nm}^I(\rr)}{\partial \R_I}
        \rho_{nm}^I  \nonumber\\
&& \quad
   - \sum_{nm} D_{nm}^I \pard{\rho  _{nm}^I}{\R_I}
   + \sum_{nm} q_{nm}   \pard{\omega_{nm}^I}{\R_I} \;,
\label{fion}
\mend
where $D_{nm}^I$ and $V\beff$ have been defined in Eqs.\ (\ref{dij})
and (\ref{veff}), respectively.  The last term of Eq.\ (\ref{fion})
gives the constraint contribution to the forces.
Since the PW basis set does not depend on atomic positions,
Pulay-type corrections \cite{Pulay} do not appear in the
expression for the forces.

\subsection{Discretization of the equation of motion 
and orthonormality constraints}
\label{evolut}

The equations of motion (\ref{ele}) and (\ref{ions}) are usually
discretized using the Verlet or the velocity-Verlet
algorithms. The following discussion, including the treatment of 
the $\R_I$-dependence of the orthonormality constraints, applies
to the Verlet algorithm when using the Fourier acceleration scheme 
of Ref.\cite{prec}. In this framework the fictitious electron mass 
is represented by an operator $\Theta$, whose matrix elements 
between PWs are given by
\begin{equation}
\Theta_{\G,\G'} = \mbox{max}\left(\mu,\mu{\hbar^2G^2\over 2m E_c}\right)
\delta_{\G,\G'},
\end{equation}
where $\G, \G'$ are the wave vector of PWs, $E_c$ is a cutoff 
(typically a few Ry) which defines the threshold for Fourier 
acceleration. The fictitious electron mass depends on $G$ as 
the kinetic energy for large $G$, it is constant for small $G$.
This scheme allows us to use larger
time steps with negligible computational overhead.

The electronic orbitals at time $t+\Delta t$ are given by:
\begin{eqnarray}
\phi_i(t+\Delta t)=2\phi_i(t)-\phi_i(t-\Delta t)
 - (\Delta t)^2\Theta^{-1}\Big[ \frac{\delta E\btot}{\delta\phi^*_i}
 - \sum_j \Lambda_{ij}(t+\Delta t)\; S(t)\phi_j(t)\Big] ,
\label{vele}
\end{eqnarray}
where $\Delta t$ is the time step, and 
$S(t)$ indicates the operator $S$ evaluated for ionic
positions $R_I(t)$. Similarly the ionic coordinates
at time $t+\Delta t$ are given by:
\begin{eqnarray}
\R_I(t+\Delta t)&=&2\R_I(t)-\R_I(t-\Delta t)\nonumber \\
&& -\frac{(\Delta t)^2}{M_I}\Big[\pard{E\btot}{\R_I}
-\sum_{ij} \Lambda_{ij}(t+\Delta t)
\langle\phi_i(t)|\pard{S(t)}{\R_I}|\phi_j(t) \rangle\Big] \;\;.
\label{vions}
\end{eqnarray}
The orthonormality conditions must be imposed at each time-step:
\begin{equation}
\langle \phi_i(t+\Delta t)|S(t+\Delta t)|\phi_j(t +\Delta t)\rangle
=\delta_{ij} \ ,
\label{abc1}
\end{equation}
leading to the following matrix equation:
\mbeg
 A + \lambda B + B^{\dagger}\lambda^{\dagger}
 + \lambda C \lambda^{\dagger} =1  \label{abc2}
\mend
where the unknown matrix $\lambda$ is related to the 
matrix of Lagrange multipliers $\Lambda$ at time
$t+\Delta t$ via 
$\lambda=(\Delta t)^2 \Lambda^*(t+\Delta t)$.
In Eq.(\ref{abc2}) the dagger indicates Hermitian conjugate
($\lambda=\lambda^{\dagger}$).
The matrices $A$, $B$, and $C$ are given by:
\mbeg
 A_{ij} &=& \bra{\bar \phi_i} S(t+\Delta t) \ket{\bar \phi_j}\nonumber
   \;\;, \\
 B_{ij} &=& \bra{\Theta^{-1} S(t)\phi_i(t)} S(t+\Delta t)
            \ket{\bar \phi_j} \nonumber
   \;\;, \\
 C_{ij} &=& \bra{\Theta^{-1} S(t) \phi_i(t)} S(t+\Delta t)
            \ket{\Theta^{-1} S(t) \phi_j(t)} \;\;,
\label{ABC}
\mend
with
\mbeg
 {\bar \phi_i }  &=& 2\phi_i(t) -\phi_i(t-\Delta t)  -
  (\Delta t)^2\Theta^{-1}\var{E\btot(t)}{\phi^*_i}
  \;\; .
\label{barphi}
\mend
The solution of Eq.\ (\ref{abc2}) in the ultrasoft PP case
is not obvious, because
Eq.\ (\ref{vions}) is not a closed expression for $\R_I(t+\Delta t)$.
The problem is that $\Lambda(t+\Delta t)$ appearing in Eq.\ (\ref{vions})
depends implicitly on $\R_I(t+\Delta t)$ through $S(t+\Delta t)$.
Consequently, it is in principle
necessary to solve iteratively for $\R_I(t+\Delta t)$ in 
Eq.\ (\ref{vions}).

A simple solution to this problem is given in Ref. \cite{CarPasquarello}. 
$\Lambda(t+\Delta t)$ is extrapolated using two previous values:
\begin{equation}
\Lambda^{(0)}_{ij}(t+\Delta t)=2\Lambda_{ij}(t)-\Lambda_{ij}(t-\Delta t).
\label{guess}
\end{equation}
Eq.\ (\ref{vions}) is used to find $\R^{(0)}_I(t+\Delta t)$,
which is correct to $O(\Delta t^4)$. From $\R^{(0)}_I(t+\Delta t)$
we can obtain a new set $\Lambda^{(1)}_{ij}(t+\Delta t)$ and 
repeat the procedure until convergence is achieved.
It turns out that in most practical applications 
the procedure converges at the very first iteration.
Thus, the operations described above are generally executed only
once per time step.

The solution of Eq.\ (\ref{abc2}) is found using a modified version
\cite{CarPasquarello,Cavazzoni} of the iterative procedure of
Ref.\ \cite{cp89}.
The matrix $B$ is decomposed into hermitian $(B_h)$ and antihermitian
$(B_a)$ parts,
\mbeg
B=B_h+B_a,
\mend
and the solution is obtained by iteration:
\mbeg
\lambda^{(n+1)}B_h+B_h\lambda^{(n+1)}=1-A-\lambda^{(n)}B_a
-B_a^{\dagger}\lambda^{(n)} -\lambda^{(n)} C\lambda^{(n)}.
\label{ortho1}
\mend
The initial guess $\lambda^{(0)}$ can be obtained from
\mbeg
\lambda^{(0)} B_h+ B_h\lambda^{(0)} =1-A .    \label{exact}
\mend
Here the $B_a$- and $C$-dependent terms are neglected because
they are of higher order in $\Delta t$ ($B_a$ vanishes
for vanishing $\Delta t$). Eqs.\ (\ref{exact}) and (\ref{ortho1})
have the same structure:
\mbeg
\lambda B_h+ B_h\lambda = X \label{ortho2}
\mend
where $X$ a Hermitian matrix.  Eq. (\ref{ortho2}) 
can be solved exactly by finding the
unitary matrix $U$ that diagonalizes $B_h$: $U^{\dagger}B_hU=D$, where
$D_{ij}=d_i\delta_{ij}$. The solution is obtained from
\mbeg
  (U^{\dagger} \lambda U)_{ij} = (U^{\dagger} X U)_{ij}/(d_i+d_j).
 \label{ortho3}
\mend
When $X=1-A$ Eq.(\ref{ortho3}) yields the starting $\lambda^{(0)}$,
while $\lambda^{(n+1)}$ is obtained from $\lambda^{(n)}$ by solving 
Eq.\ (\ref{ortho3}) with $X$ given by Eq.\ (\ref{ortho1}).
This iterative procedure usually converges in ten steps or less.

\subsection{Ultrasoft pseudopotential implementation in the serial case}

\subsubsection{Plane-wave expansion}

Let $\{\R\}$ be the translation vectors of the periodically 
repeated supercell. The corresponding reciprocal lattice 
vectors $\{\G\}$ obey the conditions $\R_i\cdot\G_j = 2\pi n$, 
with $n$ an integer number.

The KS orbitals can be expanded in a PW basis up to a 
kinetic energy cutoff $E\bcpw$:
\mbeg
  \phi_{j,{\bf k}}(\rr) = \frac{1}{\sqrt{\Omega}}
  \sum_{\G\in\{\G\bcpw\}} \phi_{j,{\bf k}}(\G)
e^{-i({\bf k}+\G){\bf \cdot}\rr},
\mend
where $\Omega$ is the volume of the cell,
$\{\G\bcpw\}$ is the set of $\G$ vectors satisfying
the condition
\mbeg \frac{\hbar^2}{2m}|{\bf k}+{\bf G}|^2 < E\bcpw,
\mend
and \k\ is the Bloch vector of the electronic states.
In crystals, one must use a grid of \k-points dense enough to
sample the Brillouin zone (the unit cell of the reciprocal lattice).
In molecules, liquids and in general if the simulation cell is large enough,
the 
Brillouin zone can be sampled using only the $\k=0$ ($\Gamma$) point.
An advantage of this choice is that the orbitals can be 
taken to be real in \rr-space. In the following we will drop
the {\bf k} vector index. 
Functions in real space and their Fourier transforms will be
denoted by the symbols, if this does not originate ambiguity.

The $\phi_j(\G)$'s are the electronic variables. 
The calculation of $H\phi_j$ and of the forces acting on
the ions are the basic ingredients of the computation.
Scalar products $\bk{\phi_j}{\beta^I_n}$ and their 
spatial derivatives are typically evaluated in \G-space.
An important advantage of working in \G-space is that
atom-centered functions like $\beta^I_n$ and $Q_{nm}^I$ 
are easily evaluated at any atomic position, for example: 
\mbeg
\beta_n^I(\G)=\beta_n(\G)e^{-i{\bf G\cdot}\R_I}.
\label{fourier}
\mend
Thus:
\mbeg
\bk{\phi_j}{\beta^I_n} = \sum_{\G\in\{\G\bcpw\}} \phi_j^*(\G) \beta_n(\G)
e^{-i{\bf G}{\bf \cdot}\R_I} \;,
\label{bec}
\mend
and
\mbeg
\bk{\phi_j}{\frac{\partial\beta^I_n}{\partial\R_I}}
= -i \sum_{\G\in\{\G\bcpw\}} {\bf G} \phi_j^*(\G) \beta_n(\G)
e^{-i{\bf G}{\bf \cdot}\R_I} \;.
\label{becdr}
\mend
The kinetic energy term is diagonal in \G-space and is easily
calculated:
\mbeg
-\left(\nabla^2\phi_j\right) (\G) = G^2 \phi_j(\G) \;.
\mend
In summary, the kinetic and nonlocal PP terms in $H\phi_j$ are
calculated in \G-space.

\subsubsection{Dual space technique}

The local potential term $V\beff\phi_j$ could be calculated in \G-space, 
but it is more convenient to use a
different (``dual space'') technique.
The idea is to switch from \G- to \rr-space, back and forth,
using FFT, and to perform the calculation in the space where
it is more convenient. The KS orbitals are first Fourier-transformed
to \rr-space; then, $(V\beff\phi_j)(\rr) = V\beff(\rr)\phi_j(\rr)$ is 
calculated in \rr-space, where $V\beff$ is diagonal; finally 
$(V\beff\phi_j)(\rr)$ is Fourier-transformed back to $(V\beff\phi_j)(\G)$.

In order to use FFT, one discretizes the \rr-space by
a uniform grid spanning the unit cell:
\begin{equation}
f(m_1,m_2,m_3) \equiv f(\rr_{m_1,m_2,m_3}), \quad 
\rr_{m_1,m_2,m_3} = m_1{\a_1\over N_1} 
                  + m_2{\a_2\over N_2}
                  + m_3{\a_3\over N_3},
\label{FFTr}
\end{equation}
where $\a_1, \a_2, \a_3$ are lattice basis vectors,
the integer index $m_1$ runs from $0$ to $N_1-1$, and 
similarly for $m_2$ and $m_3$. 
In the following we will assume
for simplicity that $N_1, N_2, N_3$ are even numbers.
The FFT maps a discrete periodic function in real space 
$f(m_1,m_2,m_3)$ into a discrete periodic function in reciprocal space 
$\widetilde f(n_1,n_2,n_3)$ (where $n_1$ runs from 
$0$ to $N_1-1$, and similarly for $n_2$ and $n_3$), and vice versa.

The link between \G-space components and FFT indices is:
\begin{equation}
\widetilde f(n_1,n_2,n_3) \equiv f(\G_{n'_1,n'_2,n'_3}), \quad
\G_{n'_1,n'_2,n'_3}=n'_1\b_1+n'_2\b_2+n'_3\b_3
\label{FFTarray}
\end{equation}
where $n_1=n'_1$ if $n'_1 \ge 0$,  
$n_1=n'_1+N_1$ if $n'_1 < 0$, and
similarly for $n_2$ and $n_3$.
The FFT dimensions $N_1, N_2, N_3$ 
must be big enough to include all non negligible Fourier 
components of the function to be transformed:
ideally the Fourier component corresponding to 
$n'_1 = N_1/2$, and similar for $n'_2$ and $n'_3$, 
should vanish. 
In the following, we will refer to the set of
indices $n_1, n_2, n_3$ and to the corresponding
Fourier components as the ``FFT grid''.

The soft part of the charge density:
$n\bsoft(\rr) = \sum_j |\phi_j(\rr)|^2$, 
contains Fourier
components up to a kinetic energy cutoff $E\bcps=4E\bcpw$.
This is evident from the formula:
\mbeg
n\bsoft(\G) = \sum_{\G'\in\{\G\bcpw\}}\sum_j \phi^*_j(\G-\G') \phi_j(\G').
\mend
In the case of NC PPs, the entire charge density is given
by $n\bsoft(\rr)$.

$V\beff$ should be expanded up to the same $E\bcps$ cutoff since 
all the Fourier components of
$V\beff\phi_j$ up to $E\bcpw$ are required.
Let us call $\{\G\bcps\}$ the set of \G-vectors such that
\mbeg
\frac{\hbar}{2m}G^2 < E\bcps.
\label{FFTsoft}
\mend
The soft part of the charge density is conveniently calculated
in \rr-space, by Fourier-transforming $\phi_j(\G)$ into
$\phi_j(\rr)$ and summing over the occupied states.

The exchange-correlation potential $\mu\bxc(\rr)$, 
Eq. (\ref{vxc}), is a function of the local charge density 
and -- for gradient-corrected functionals -- of its gradient 
at point \rr:
\mbeg
\mu\bxc(\rr) = V\bxc(n(\rr), |\nabla n(\rr)|)\;.
\label{Vxc}
\mend
The gradient $\nabla n(\rr)$ is conveniently calculated
from the charge density in \G-space, using 
$(\nabla n)(\G)= -i\G n(\G)$.
The Hartree potential $V\bh(\rr)$, Eq. (\ref{vh}), 
is also conveniently calculated in \G-space:
\mbeg
V\bh(\G) = {4\pi\over \Omega} \frac{n(\G)^*}{ G^2} \;.
\label{VH}
\mend
Thus in the NC-PP case, a single FFT grid, large enough to 
accommodate the $\{\G\bcps\}$ set, can be used for orbitals, 
charge density, and potential.

The use of FFT is mathematically equivalent to a pure \G-space description
(we neglect here a small inconsistency in exchange-correlation potential 
and energy density, due to the presence of a small amount of components 
beyond the $\{\G\bcps\}$ set). This has important 
consequences: working in \G-space means that translational 
invariance is exactly conserved
and that forces are analytical derivatives of 
the energy (apart from the effect of the small inconsistency mentioned above).
Forces that are analytical derivatives of the energy ensure that the
constant of motion (i.e., the sum of kinetic and potential energy 
of the ions in Newtonian dynamics) is conserved during the evolution.

\subsubsection{Double-grid technique}

Let us focus on US PPs.
In \G-space the charge density is:
\mbeg
 n(\G) =  n\bsoft(\G) +
 \sum_{i,nm,I} Q^I_{mn}(\G) \bk{\phi_i}{\beta^I_n}
 \bk{\beta^I_m}{\phi_i} \;\;.
\label{gdens}
\mend
If $E\bcpw$ is the cutoff for the KS orbitals,
the cutoff for the soft part of the charge density 
is $E\bcps=4E\bcpw$. The augmentation term often requires
a cutoff higher than $E\bcps$, and as a consequence a
larger set of \G-vectors. Let us call $\{\G\bcpn\}$
the set of \G-vectors that are needed for the augmented part:
\mbeg
\frac{\hbar^2}{2m}G^2<E\bcpn.
\label{FFThard}
\mend
In typical situations, using pseudized augmented charges,
$E\bcpn$ ranges from $E\bcps$ to 
$\sim 2-3E\bcps$.

The same FFT grid could be used both for the augmented
charge density and for KS orbitals. This however would imply using an 
oversized FFT grid in the most expensive part of the calculation,
dramatically increasing computer time.
A better solution is to introduce
two FFT grids:
\begin{itemize}
\item
a coarser grid (in \rr-space) for the KS orbitals 
and the soft part of the charge density. The
FFT dimensions  $N_1, N_2, N_3$ of this grid are big
enough to accommodate all \G-vectors in $\{\G\bcps\}$
\item
a denser grid (in \rr-space) for the total charge density
and the exchange-correlation and Hartree potentials. The
FFT dimensions $M_1\ge N_1, M_2\ge N_2, M_3\ge N_3$ of this grid are big
enough to accommodate all \G-vectors in $\{\G\bcpn\}$. 
\end{itemize}

In this framework, the soft part of the electron density $n\bsoft$, 
is calculated in \rr-space using FFTs on the coarse grid and transformed 
in \G-space using a coarse-grid FFT on the $\{\G\bcps\}$ grid.
The augmented charge density is calculated in \G-space on the
$\{\G\bcpn\}$ grid, using Eq. (\ref{gdens}) as described in the
next section.
$n(\G)$ is used to evaluate the Hartree potential, Eq.(\ref{VH}).
Then $n(\G)$ is Fourier-transformed in \rr-space on the dense grid,
where the exchange-correlation potential, Eq.(\ref{Vxc}), is
evaluated.

In real space, the two grids are not necessarily commensurate.
Whenever the need arises to go from the coarse to the dense grid,
or vice versa, this is done in \G-space. For instance, the 
potential $V\beff$, Eq.\ (\ref{veff}), is needed both
on the dense grid
to calculate quantities such as the $D_{nm}^I$, Eq.\ (\ref{dij}), 
and
on the coarse grid to calculate $V\beff\phi_j$, Eq.\ (\ref{ks}). 
The connection between
the two grids occurs in \G-space, where Fourier filtering is performed:
$V\beff$ is first transformed in
\G-space on the dense grid, then transferred to the
coarse \G-space grid by eliminating components incompatible with $E\bcps$,
and then back-transformed in \rr-space using a coarse-grid FFT.

We remark that for each time-step only a few dense-grid FFT
are performed, while the number of necessary coarse-grid FFTs 
is much larger, proportional to the number of KS states $N\bks$.

\subsubsection{Augmentation boxes}
\label{serialbox}

Let us consider the augmentation functions $Q_{nm}$, which
appear in the calculation of the electron density, Eq.\ (\ref{gdens}), 
in the calculation of $D_{nm}^I$, Eq.\ (\ref{dij}), and in the
integrals involving $\partial Q^I_{nm}/\partial{\bf R}_I$ 
needed to compute the ionic forces, Eq.\ (\ref{fion}).
The calculation of the $Q_{nm}$ in \G-space has a large
computational cost because the cutoff for the $Q_{nm}$
is the large cutoff $E\bcpn$. The computational cost can be
significantly reduced if we take advantage of the localization
of the $Q_{nm}$ in the core region.

We call ``augmentation box'' a fraction of the supercell,
containing a small portion of the dense grid in real space.
An augmentation box is defined only for atoms described by
US PPs.
The augmentation box for atom $I$ is centered at the point 
of the dense grid that is closer to the position $\R_I$.
During a MD run, the center of the $I-$th augmentation box
makes discontinuous jumps to one of the neighboring grid points
whenever the position vector $\R_I$ gets closer to such grid point. 
In a MD run, the augmentation box must always contain completely 
the augmented charge belonging to the $I-$th atom; otherwise, the 
augmentation box must be as small as possible. Augmentation
boxes of different sizes for different atoms could in principle
be used, but in our implementation the same box size is chosen for 
all the atoms. Thus the atomic species having the less localized
augmented charge determines the size of all the augmentation
boxes.

The volume of the augmentation box is much smaller than the volume 
of the supercell. The number of \G-vectors in the reciprocal space
of the augmentation box is smaller than the number of \G-vectors 
in the dense grid by the ratio of the volumes of the augmentation 
box and of the supercell. As a consequence, the cost of calculations
on the augmentation boxes increases linearly with the number of
atoms described by US PPs. 

Augmentation boxes are used twice in the calculation:
\begin{itemize}
\item [(i)] to construct the augmented charge density, Eq.\ (\ref{dens}),
\item [(ii)] to calculate the self-consistent contribution to
     the coefficients of the nonlocal PP,  Eq.\ (\ref{dij}).
\end{itemize}

In case (i), the augmented charge is conveniently 
calculated in \G-space, following 
Ref.\cite{CarPasquarello},
and Fourier-transformed in \rr-space. All these calculations are done
on the augmentation box grid. Then
the calculated contribution at each \rr-point of the augmentation box
grid is added to the charge density at the same point in the dense grid.
In case (ii), it is convenient to calculate $D_{nm}^I$ as follows:
for every US atom,
take the Fourier transform of $V\beff(\rr)$ on the corresponding 
augmentation box grid and evaluate the 
integral of Eq.\ (\ref{dij}) in \G-space.

\section{Parallel Ultrasoft Pseudopotential Implementation}

Various parallelization strategies for PW-PP calculations 
have been described in the literature. A strategy that ensures 
excellent scalability in terms of both computer time and memory 
consists 
in distributing the PW basis set and the FFT grid points in real 
and reciprocal space across processors. A crucial issue
for the success of this approach is the FFT algorithm,
which must be capable of performing three-dimensional FFT 
on data shared across
different processors with good load balancing \cite{ParaFFT}. 
This algorithm can be generalized to the US case as
described in the following subsection.

\subsection{Parallel FFT in the US case}

Partitioning a real-space FFT grid across processors is
straightforward. The FFT grid, Eq.(\ref{FFTr}), is 
subdivided in a number
of slices equal to the number of processors, so that 
each processor can take care of a different slice.
The slices are cut along planes orthogonal to a chosen
crystallographic direction. We label the crystallographic 
directions by 1,2,3.
For instance, let us consider a FFT grid with $N_3$ planes 
along direction 3, which is distributed across $N_p$ 
processors. If $N_p$ is a divisor of $N_3$, good load 
balancing is achieved if each slice contains the same
number $(N_3/N_p)$ of planes. Processor $p$ contains
planes with $m_3$ values such that:
$(p-1)(N_3/N_p)\le m_3\le p(N_3/N_p)-1$.
If $N_p$ is not a divisor of $N_3$, all the slices cannot 
be equal. In this case their dimension is chosen in such
a way as to minimize load imbalance. If $N_p$ exceeds 
the number of planes $N_3$, this strategy has to be 
refined.

The partition of the \G-space grid is more involved. 
The Fourier components of the quantities of interest
(e.g. the orbitals, the charge density, etc.)
are stored as vectors (one-dimensional arrays): 
$f(i) \equiv f(\G_i)$, where the index $i$ spans
one of the three sets of \G-vectors defined above,
namely the set $\{\G\bcpw\}$, the set $\{\G\bcps\}$,
and the set $\{\G\bcpn\}$.
When a FFT is needed, the Fourier components have to be 
transferred to one of the two grids (three-dimensional arrays), 
defined by Eq.(\ref{FFTarray}). The two grids are either the
coarse grid, with dimensions $N_1, N_2, N_3$, or the dense
grid, with dimensions $M_1, M_2, M_3$. The Fourier components 
must be evenly distributed across processors 
in order to achieve optimal load balancing for operations
like scalar products. At the same time, their distribution
across processors should achieve good load balancing in 
the FFTs {\em and} minimize the amount of data communication 
needed to perform the FFTs.

For each pair $n'_1, n'_2$ in Eq.(\ref{FFTarray}) we define a 
``column'' in \G-space,
including all $\G_{n'_1,n'_2,n'_3}$ with $-M_3/2 \le n'_3 \le M_3/2$.
Since the KS orbitals have nonzero Fourier component only 
for \G-vectors belonging to the set $\{\G\bcpw\}$, only a subset 
of all the columns contributes to a one-dimensional FFT of
a KS orbital in the direction $3$. We call these columns 
``active columns'' for the set $\{\G\bcpw\}$. 
In general, the number of nonzero Fourier components is 
different for each active column.
Ideally, we would like to distribute the active columns across 
the processors, so that each processor receives the same number
of active columns and the same number of Fourier components. 
Although not possible in general, this can be achieved 
to a good extent with a simple algorithm\cite{SdG}:
1) create a list of columns, ordered by decreasing number of nonzero 
Fourier components;
2) assign the column to the processors, following the order in the 
list;
3) when all the processors contain at least one column, assign
the following column in the list to the processor with the
smallest number of nonzero Fourier components.
This algorithm works nicely when the number of columns
per processor is large enough.

After assigning to the processors all the columns that are
active for the set $\{\G\bcpw\}$, we distribute
across the processors the remaining columns, that are active
for the set $\{\G\bcps\}$, using the same algorithm.
Finally we distribute across the processors the remaining columns, 
that are active for the set $\{\G\bcpn\}$, again
using the same algorithm. The remaining columns are not active
for any set of \G-vectors and play no role.

After distributing all the columns across the processors,
a one-dimensional FFT along direction 3 is done on 
local data (on a single processor). However, the data
on the planes orthogonal to direction 3 are distributed
across the processors. In order to perform FFTs in each
of these planes, the corresponding data must be made
local to a processor. This is achieved by a parallel 
transpose operation, performed with a single call to 
the appropriate MPI\cite{MPI} library routine.
Two-dimensional FFTs can then be performed 
on the planes, with each processor operating on
local data. Nonzero contributions are present only 
for $(n'_1,n'_2)$ pairs corresponding to active columns. 
This fact can be exploited to reduce the number of FFT 
operations, by performing only the FFTs along direction 
1 (or 2) that include nonzero contributions. 
The strategy for parallel three-dimensional FFT that we have
presented requires the number of processors $N_p$ be smaller 
than or equal to the number of planes $N_3$.
The FFT from \rr- to \G-space uses the same algorithm in
reversed sequence. 

In calculations using only the $\Gamma$ point ($\k=0$),
the KS orbitals can be chosen to be real functions in \rr-space, 
so that $\phi(\G)=\phi^*(-\G)$. This allows us to store 
only half of the Fourier components.
Moreover, two real FFTs can be performed as a single complex FFT.
To this end the auxiliary function $\Phi$ is introduced:
\mbeg
\Phi(\rr) = \phi_j(\rr)+ i \phi_{j+1}(\rr)
\mend
whose Fourier transform $\Phi(\G)$ yields
\mbeg
   \phi_j    (\G) & = & {\Phi(\G) + \Phi^*(\G)\over 2} \\
   \phi_{j+1}(\G) & = & {\Phi(\G) - \Phi^*(\G)\over 2i}.
\mend
A side effect on parallelization is that $\G$ and $-\G$ must
reside on the same processor. As a consequence, pairs of columns
with $\G_{n'_1,n'_2,n'_3}$ and $\G_{-n'_1,-n'_2,n'_3}$ 
(with the exception of the case $n'_1=n'_2=0$), 
must be assigned to the same processor.

\subsection{Scalar products}

All scalar products $\langle f| g\rangle = \sum_i f^*_i g_i, i=1,n$
where $i$ runs on a distributed grid, can be calculated
by calling standard optimized library routines (like BLAS from
NetLib\cite{NetLib}) on each processor, and 
subsequently by summing the partial results of all processors, 
using a call to standard MPI\cite{MPI} libraries. Scalar products 
between vectors for which only half of the Fourier components 
are stored require a special treatment.
Let $n_p$ be the number of Fourier components stored on processor $p$.
The contribution of this processor to the scalar product is
$\langle f| g\rangle_p =2 \sum_{i=1,n_p} f^*_i g_i$
if the $\G=0$ components are not within the set of $n_p$ 
components. If instead the $\G=0$ components, identified by $i=1$, 
are stored on processor $p$, the contribution of processor $p$ 
to the scalar product is
$\langle f| g\rangle_p = f_1 g_1 + 2 \sum_{i=2,n_p} f^*_i g_i$.

\subsection{Iterative Orthonormalization}

The scalar products in the matrix elements, Eq. (\ref{ABC}), 
needed to compute the Lagrange multipliers are calculated
in parallel, following the procedure of the previous subsection.
In the present
implementation, the solution of the matrix equation (\ref{ortho2}),
involving square matrices of dimension equal to 
the number $N\bks$ of KS orbitals, is not
parallelized but replicated on all the processors.
Usually the time spent in the non parallelized part of the iterative 
orthonormalization is only a small fraction of the total time
of the calculation.

To efficiently perform calculations on very large systems, using a
large number of processors,
the solution of Eq. (\ref{ortho2})
should also be parallelized.
The time consuming steps are
matrix-matrix multiplication and the diagonalization of the
$B$ matrix. Both calculations require ${\cal O}(N\bks^3)$ floating-point 
operations. A convenient parallelization approach
is described in Ref.\cite{Cavazzoni}.

\subsection{Augmentation boxes}

The parallelization of the calculations
performed on the augmentation boxes is not obvious
for two reasons:
1) each augmentation box has a grid which is 
a portion of the dense grid and is distributed across
processors;
2) the boxes follow the atoms in the MD evolution,
causing the portion of the dense grid to change with time.
In the present implementation, we deal with these difficulties
as follows. We keep on all processors a copy
of all the quantities defined on the augmentation boxes.
Calculations on the grid of a given augmentation-box 
are performed only in the processors that contain
at least a fraction of the given augmentation-box grid.
This causes some replication of the calculations.
FFTs on the augmentation box grid are performed locally 
on each processor. In order to reduce the amount of
replication, in the FFTs from \G- to \rr-space,
the two-dimensional FFTs along planes orthogonal to direction 3
are performed only in the planes belonging to the slice of
the dense grid that is local to a given processor. 
No communication is needed to copy the augmented charge 
in \rr-space from the augmentation-box grid to the dense grid
(see Sec. \ref{serialbox}).
In the calculation of $D_{nm}^I$, we evaluate $Q_{nm}^I$ in \G-space, 
transform it in \rr-space using augmented-box FFT, evaluate the 
integral of Eq.\ (\ref{dij}) in \rr-space,and  sum the final result 
over all processors. This approach keeps communications to a minimum,
at the expense of a number of augmentation-box FFTs larger than
in the serial case.

Augmentation-box grid related calculations constitute a very small part of the 
overall computational cost, both in computer time and in memory. Therefore
the simple approach that we have presented is convenient even if some calculations
are replicated on few processors and the load balance is not optimal.

\section{Test case: iron porphyrins}

We report here a comparison of computer performances for US and 
standard PPs in CP calculations. Our test systems--prototypes 
of systems containing the iron-porphyrin motif--are a reduced 
and an extended model of the active site of myoglobin.

\subsection{Models and computational details}

The reduced model is composed of an iron-porphyrin-imidazole complex,
already investigated using the CP method by Rovira et al. \cite{Rovira};
the metallic pentacoordinated center is bound to the four planar
porphyrin nitrogens, 
with the imidazole nitrogen occupying one of the axial sites,
binding approximately orthogonal to the porphyrin plane, see Figure 1.
The chemical formula is [FeN$_6$C$_{23}$H$_{16}$]. A simple cubic cell of
size 15.875~\AA, containing a total of 46 atoms and 154 electrons, is used.
For the reduced model, we compare both spin-restricted and
unrestricted (S=2) calculations. 

\begin{figure}
\begin{center}
\epsfxsize=8.5cm
\epsfbox{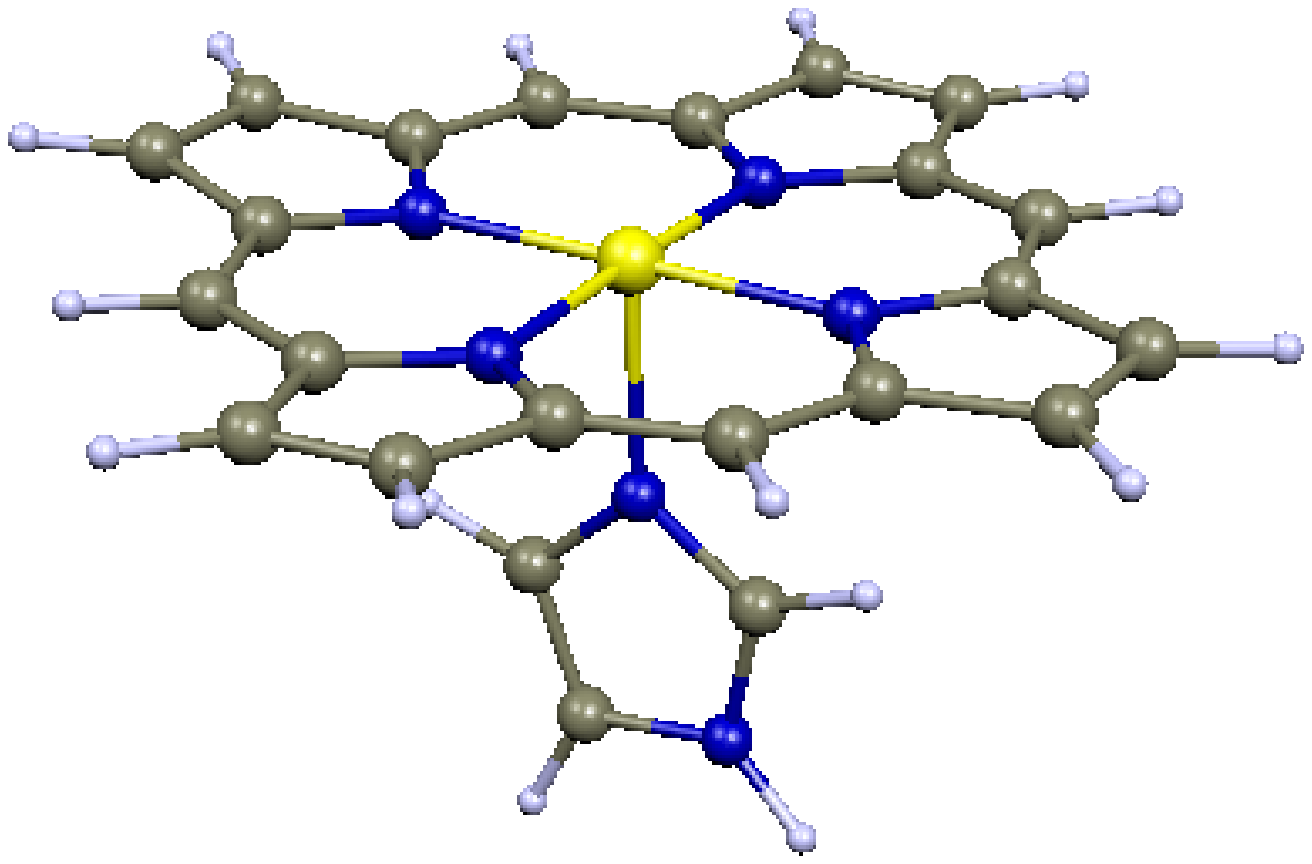}
\end{center}
Figure 1. Reduced model: the iron-porphyrin-imidazole complex.
Yellow: Fe. Dark gray: C. Blue: N. Light gray: H.
\end{figure}

The extended model is composed by a large portion of the myoglobin
active site, defined by the full heme group (same coordination as for
the reduced model) plus the 13 surrounding residues 
which were comprised within a sphere of 8~\AA\ radius
centered on the iron atom, see Figure 2.
The initial geometry has been taken from the X-ray experimental 
structure of the O$_2$-myoglobin complex \cite{X-ray}
and the included residues have been terminated 
by NH$_2$ groups,
resulting in a total of 332 atoms and 902 electrons.
The chemical formula is [FeO$_{19}$N$_{35}$C$_{106}$H$_{173}$].
A simple cubic cell of size 25.4~\AA\ has been used,
ensuring a minimum separation of 5~\AA\ between 
periodic replicas. 
For the extended model we discuss only the performances
of the more computationally demanding spin-unrestricted (S=2) calculations.
\begin{figure}
\begin{center}
\epsfxsize=16.5cm
\epsfbox{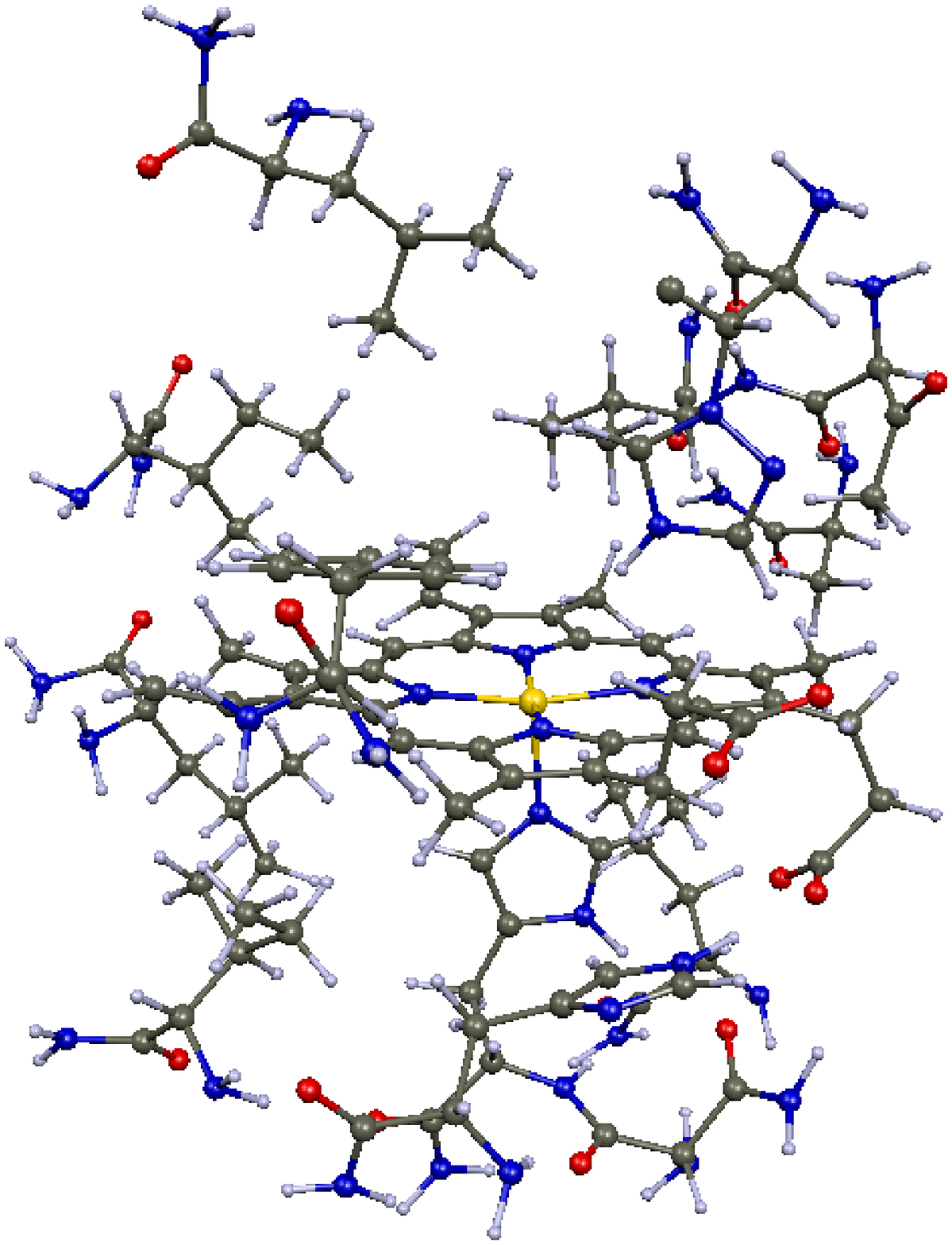}
\end{center}
Figure 2. Extended model of the myoglobin active site.
Red: O, other colors as in Figure 1.
\end{figure}

For a correct comparison of performances we need to compare data
of similar quality, in terms of accuracy of chemical properties, 
obtained with algorithms of comparable quality in terms of 
serial speed and parallel speedup. In order to satisfy the first 
requirement, we need to determine a set of cutoffs for the US and 
standard calculations yielding comparable structural properties. 
To this end we compared the optimized geometry of the triplet ground
state for the reduced model from our US calculations and from 
published NC results, performed at $E\bcpw=70$ Ry
\cite{Rovira}.
Our geometrical parameters are converged to the same extent as those
in Ref. \cite{Rovira} at $E\bcpw=25$ Ry and $E\bcpn=200$ Ry; 
in particular the critical Fe-N distances in the porphyrin and in
the imidazole are found to be 2.00 and 2.12~\AA\ respectively, 
vs. 2.00 and 2.14~\AA\ of Ref. \cite{Rovira}, 
with the out-of-plane displacement of the iron
atom computed to be 0.14 vs. 0.15~\AA.
The overall agreement between the two sets of results
is excellent, and the residual difference can be attributed 
to the different functionals used: BP86 \cite{BP86} in
Ref.\cite{Rovira}, and PW91 \cite{PW91} in our calculations.
We estimate that we can safely compare US-PP calculations
performed at $E\bcpw=25$ Ry to standard calculations at 
$E\bcpw=70$ Ry. For a fair comparison we use the same 
cutoff for the charge density in the US case as in the
standard case ($E\bcpn=280$ Ry). 
We also performed calculations at $E\bcpw=35$ Ry
for the US case, at $E\bcpw=100$ Ry for the standard case
($E\bcpn=400$ Ry in both cases). 

In order to compare algorithms of similar quality, 
all calculations were performed using the same 
code \cite{CPV} (standard PPs are just a special case of US PPs).
The PPs used in the standard case tests were 
generated using the technique of Troullier and Martins\cite{tm}.
In the US case, we use US PPs for all atoms, including H.
The PW91 functional\cite{PW91} is used in all calculations.

\subsection{Results}

The results for the reduced models were obtained on a
32-node IBM SP3 ($4\times375$MHz power3 processors per node),
while the extended model calculations were performed on a 64-processor
SGI Origin ($64\times300$MHz RISC 12000 processors), 
both at the Keck Materials Science Laboratory, Princeton University.

The reported execution times are an average over 20 time steps 
of the measured wall time (the sum of CPU and system time,
differing by only a few percent from pure CPU time).

\begin{table}[ht]
\caption{Performances of the US and NC calculations, for the spin-restricted case.
For US PPs: $E\bcpw=25$ Ry, $E\bcpn=280$ Ry, FFT grid size: 160,96,16 
for the dense, coarse, and augmentation box grids, respectively. 
For NC PPs: $E\bcpw=70$ Ry, $E\bcpn=280$ Ry, FFT grid size: 160.
$N_p$ is the number of processors;
$M_r$ is an estimate of the RAM needed per processor in Mb;
$T_e$ is the execution time per electronic
time step (at fixed atoms), in s;
$T_i$ is the same as $T_e$ per CP time step (atoms moving), in s.
}
\vspace{1cm}
\begin{tabular}{r|rrr|rrr}
      & \multicolumn {3}{c|} {US} &\multicolumn{3}{c} {NC} \\
\hline
$N_p$ & $M_r$ & $T_e$ & $T_i$ & $M_r$  & $T_e$  & $T_i$ \\
\hline         	              
  4   & 190   & 49.5  & 56.2 & 357   &136.1  &159.7 \\
  8   & 100   & 23.6  & 26.1 & 139   & 44.6  & 49.8 \\
 16   &  57   & 10.7  & 12.1 & 77   & 21.0  & 22.9
\end{tabular}
\end{table}

The parallelization performances of the US vs. NC-PPs implementation for the
spin-restricted case of the reduced model are shown in Table 1.
The execution times for a CP step (including calculation of forces and
time evolution of atomic positions) are about 15 \%\ larger than for
a purely electronic step (orbital time evolution only), as 
expected.
The small superlinear speedup observed both for US and NC PPs 
is a consequence of caching: since the 
memory per processor decreases almost linearly with the number 
of processors, better caching can be achieved with an increased 
number of processors,
thus increasing the serial speed of the code.
It is worth noting that 
US calculations are faster by a factor $\sim 2.5$ with respect to the NC case
and require half RAM memory and $1/4$ disk space with respect to
standard calculations.

The performances of US vs. NC-PPs calculations at higher cutoff are shown in
Table 2. The number of PWs is approximately $(35/25)^{3/2}\simeq 1.65$
times larger than in the preceding case. Execution times should 
theoretically be proportional to the same factor. The factor is 
actually somewhat larger ($\simeq 1.9$), but the cache effects 
mentioned above and the effect of the discreteness of the FFT 
grid explain the difference. Again,
US calculations are faster by a factor $\sim 2.5$ with respect to the NC case
and require half RAM memory with respect to
standard calculations.

In Table 3 we report spin-unrestricted results, showing an approximate
doubling of execution time and of memory requirements, in line with
expectations.

\begin{table}[ht]
\caption{Performances of the US calculations, spin-restricted case
at higher cutoff.
For US PPs: $E\bcpw=35$ Ry, $E\bcpn=400$ Ry, FFT grid size: 192,120,20
for the dense, coarse, and augmentation box grids, respectively.
For NC PPs: $E\bcpw=100$ Ry, $E\bcpn=400$ Ry, FFT grid size: 192.
The meaning of
the various columns is the same as in Table 1.
}
\vspace{1cm}
\begin{tabular}{r|rrr|rrr}
      & \multicolumn {3}{c|} {US} &\multicolumn{3}{c} {NC} \\
\hline
$N_p$ & $M_r$ & $T_e$ & $T_i$ & $M_r$  & $T_e$  & $T_i$ \\
\hline
  8   & 167   & 44.7  & 50.1 & 307   & 118.7 &143.7 \\
 16   &  93   & 20.9  & 23.0 & 157   &  50.5 & 57.2
\end{tabular}
\end{table}

\begin{table}[ht]
\caption{Performances of the US calculations: spin-unrestricted
calculations, same cutoffs as in Table 1.}
\vspace{1cm}
\begin{tabular}{r|rrr}
$N_p$ & $M_r$ & $T_e$ & $T_i$ \\
\hline         	              
  4   & 263   & 92.5  &104.5 \\   
  8   & 139   & 44.6  & 49.8 \\
 16   &  77   & 21.0  & 22.9
\end{tabular}
\end{table}
\begin{table}[ht]
\caption{Performances of the US calculations: spin-unrestricted
calculations for the extended model, $E\bcpw=25$ Ry,
 $E\bcpn=200$ Ry, FFT grid size: 224, 160, 20
for the dense, coarse, and augmentation box grids, respectively.}
\vspace{1cm}
\begin{tabular}{r|rrr}
$N_p$ & $M_r$ & $T_e$ & $T_i$ \\
\hline      		              
  16   & 1067 & 3011  & 3764 \\
  32   &  656 & 1407  & 1690 \\   
  48   &  441 &  992  & 1190 \\
\end{tabular}
\end{table}
\begin{figure}
\begin{center}
\epsfxsize=16.5cm
\epsfbox{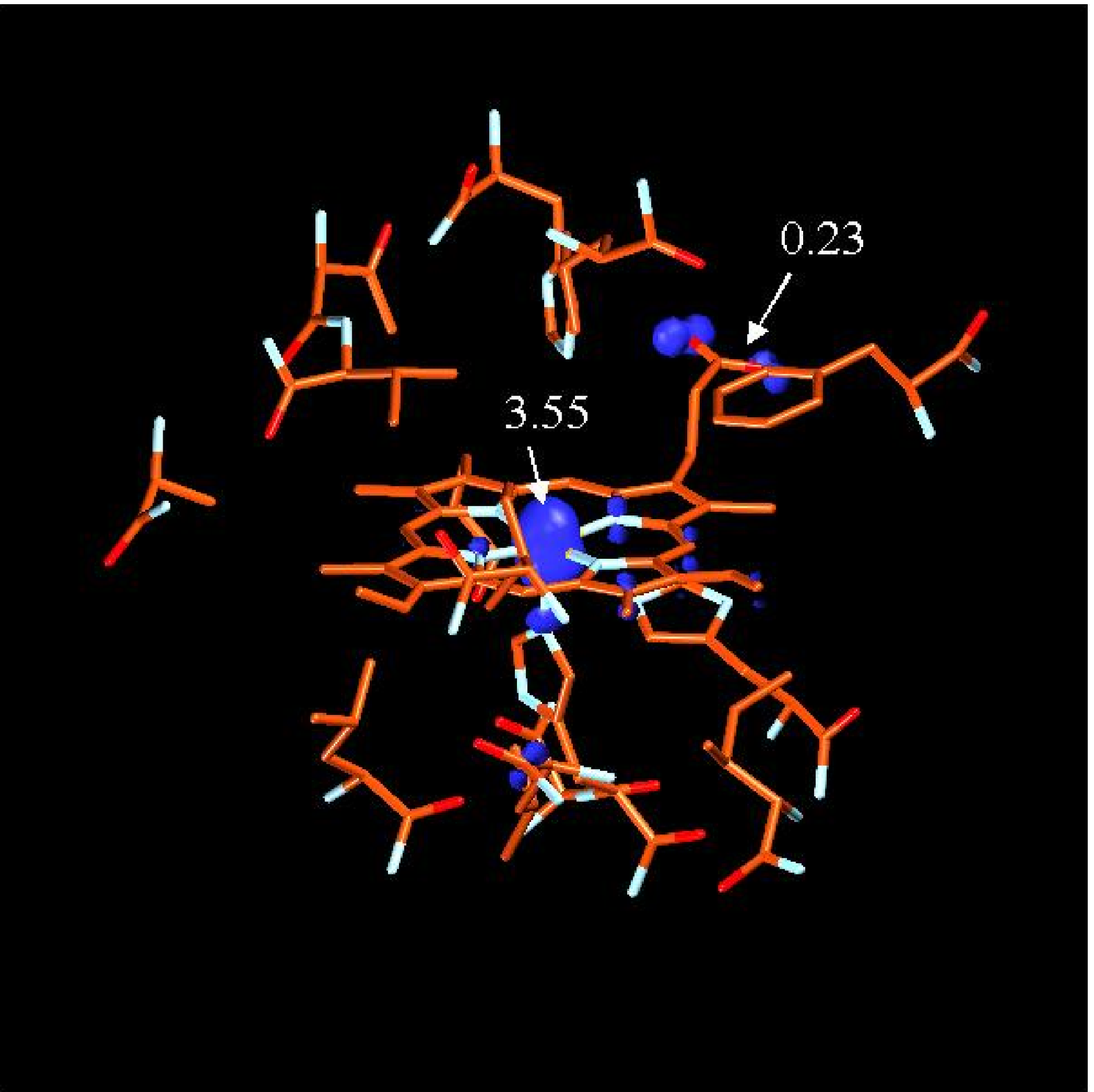}
\end{center}
Figure 3. Isodensity contour plot (contour value 0.02) and selected
integrated values of the
spin density for the quintet state
of myoglobin extended model.
Hydrogen atoms have been omitted for clarity.
\end{figure}

Table 4 contains the performances of US calculations on the quintet
state ($S=2$), which is experimentally known
to be the ground spin state in myoglobin,
of our extended model of myoglobin at $E\bcpw=25$ Ry 
and $E\bcpn=200$ Ry. 
The scaling with the number of processors is excellent in this case
too, with a slight superlinear scaling up to 32 processors. 
The ratio between execution times for electronic and CP
time-steps is almost the same in the extended and in the reduced models.
A geometry optimization time step requires less than 20 minutes in the 
spin-unrestricted case on 48 processors, with a total RAM usage of 
$\sim 21$ Gb. 
The total size of the file containing the KS orbitals is 6.7 Gb.
In this case no standard calculation was attempted;
indeed, an extrapolation from the results of Tables 1, 3 and 4
points to a total memory requirement of more than 40 Gb.

A typical local geometry optimization requires $\sim 250$
time-steps. An execution time of 20 minutes per time step thus
translates into less than 4 days for the optimization to
complete. This is a perfectly feasible calculation, not even
requiring a state-of-the art massive parallel computer.
On the other hand, a typical MD run
requires no less than 10000 time-steps, corresponding in the present case to a 
few ps of simulation time. A true dynamical simulation would 
therefore become accessible on a state-of-the art massive parallel computer.

The relevance of simulating extended portions of myoglobin active site
can be understood by considering the spin density distribution of the
quintet spin state of the extended model, reported in Figure 3
together with selected integrated spin density values. 
The spin density is
mainly localized on the iron atom (ca. 88 $\%$), 
even though a sizable contribution (ca. 12 $\%$) is computed to be
delocalized over the rest of the system, with largest contributions
arising from the propionate groups bound to the porphyrin ring. 
This finding is of particular interest,
considering that CO rebinding in myoglobin
has been recently related to a spin crossover from the quintet spin state,
corresponding to unbound CO + myoglobin,
to the singlet spin state characterizing the bound configuration of
the CO-myoglobin complex. \cite{Harvey} 
Since explicit inclusion of the protein environment alters the spin
distribution of the quintet state in our extended model, 
an effect on the relative energy of the different spin states can be expected.

\section{Accuracy of Periodic Boundary Conditions}

The use of PBC to describe molecules is perfectly appropriate
for neutral molecules with small dipole/quadrupole moments, provided
that the chosen supercell is large enough to minimize the spurious
interactions between periodic replicas. This goal can usually be
reached with supercells that leave a few \AA\ of empty space between 
periodic replicas.

Charged molecules should be described by charged supercells,
but these have infinite electrostatic energy. A finite energy 
is obtained by setting to zero the divergent $G=0$ contributions
to the energy, as if the system were neutral. This is equivalent 
to adding a neutralizing background. Energies obtained in this way 
will be referred to as ``uncorrected''. The direct comparison
of uncorrected energies between different charge states is
usually meaningless, because the error induced by PBC is large 
in this case. The long-range character of Coulomb interactions
would require unpractically large supercells.
Uncorrected energies may be affected by a large error
also in molecules with large dipole/quadrupole moments.

Several techniques have been devised to overcome such limitation. 
The Hockney technique\cite{Hockney} yields an exact treatment 
of charged species using PWs without imposing PBC. This is 
achieved by cutting the Coulomb potential in real space
beyond a suitably chosen cutoff that excludes all spurious
interactions between periodic replica, still taking into
account intramolecular interactions.
This technique is rather expensive, since it requires the 
definition of an enlarged FFT grid for the Coulomb potential.
A similar technique\cite{Martyna} where the cutoff acts in
reciprocal space allows for faster execution with minor loss
of accuracy.

A much simpler and approximate technique, due to Makov and 
Payne (MP) \cite{MakovPayne}, consists in calculating the 
leading electrostatic correction terms
and removing them from the uncorrected total energy. 
The MP correction is performed {\em a posteriori} on the
energy only: the effect of the charge on the potential 
and on atomic forces is therefore neglected. 

The Makov-Payne corrected energy $E_{MP}$ for 
a cubic supercell has the form:
\begin{equation}
E_{MP} = E + {q^2\alpha\over 2L} -{2\pi qQ\over 3L^3}
\label{MP}
\end{equation}
where $E$ is the uncorrected energy, $q$ is the net charge,
$Q$ is the quadrupole moment, $\alpha$ is the Madelung constant,
$L$ is the supercell side. This is Eq.(15) of Ref.\cite{MakovPayne}
with the correct sign.
When applying Eq. (\ref{MP}), the origin has to be translated so that the 
dipole moment vanishes \cite{MakovPayne}.
Therefore, the calculation of the Makov-Payne correction is straightforward
since it requires only calculation of the dipole and quadrupole moments.

\subsection{Models and computational details}

We want to verify the ability of PW-PP calculations with
the MP correction to reproduce the
electronic and structural properties of highly charged species.
We compute the energy difference between two {\em meso}-substituted Mn(V) 
porphyrins: the oxo-aquo-Mn(V)TM-2-Pyridyl (2-Pyp) 
and -Mn(V)TM-4-Pyridyl (4-Pyp) porphyrins, {\bf I}
(see Figure 4),
and between the corresponding oxo-hydroxo species, {\bf II},
in which the axial water
molecule has been replaced by a OH$^-$ ligand.
We compare our results to calculations employing 
a localized basis set of Slater type orbitals (STO).
\begin{figure}
\begin{center}
\epsfxsize=16.5cm
\epsfysize=14.5cm
\epsfbox{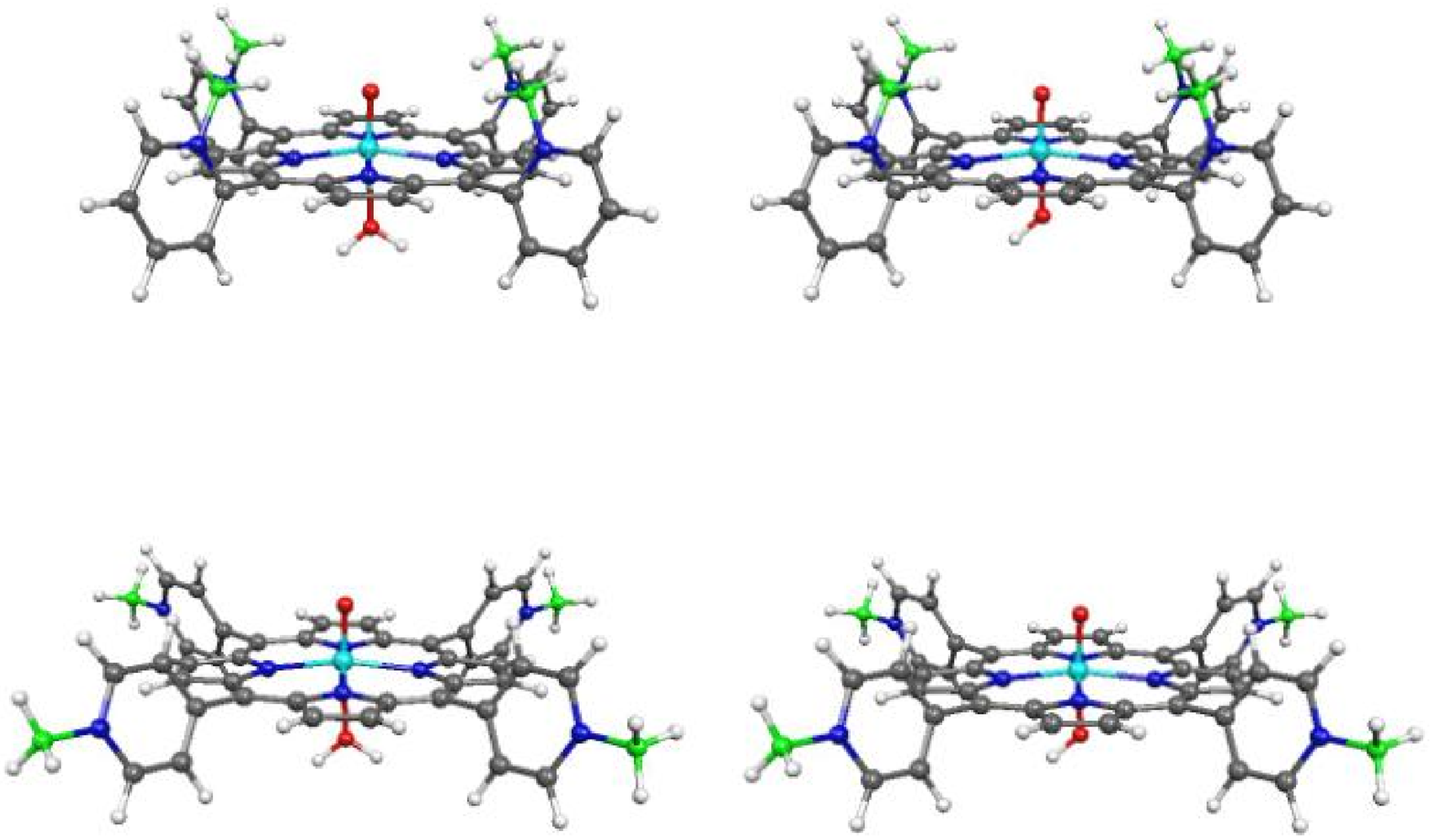}
\end{center}
Figure 4. 2-Pyp (upper panel) and 4-Pyp (lower panel) isomers of the
oxo-aquo (left panel) and oxo-hydroxo (right panel) Mn(V) porphyrins.
Mn is the light blue atom, the green atom signals the C 
of the methyl group
bound to a N in the aromatic ring, whose position differs in the
two isomers. Other colors are as in Figure 2.
\end{figure}
The oxo-aquo and oxo-hydroxo porphyrins, recently experimentally 
characterized as mimics of the halide oxidation
reaction catalyzed by haloperoxidases\cite{Groves1,Groves2},
have a charge $+5$ and $+4$, 
respectively, with 4 positive charges approximately localized on 
the aryl nitrogens and, in the case of the oxo-aquo species,
the residual positive charge located at the metal center;
the two isomeric porphyrins differ, both in the oxo-aquo and oxo-hydroxo form,
for the position of the methyl-substituted nitrogen in the aromatic
ring attached  
to the {\em meso} porphyrin carbons, which should lead to a
considerable increase  
of the quadrupole moment from the 2-Pyp to the 4-Pyp isomer.
Due to charge differences between oxo-aquo and oxo-hydroxo species,
to the high total charge and to the large expected difference in 
the quadrupole moment between the 2-Pyp and 4-Pyp porphyrins,
we believe that the calculation of the relative energies of the two 
isomeric porphyrins represents a severe test for PW calculations within PBC.

We consider a reduced model of the real systems, 
in which the methyl groups bound the the aryl nitrogens are replaced
by hydrogens. 
The US-PP results are compared to those obtained by using STOs 
in the frozen core approximation.
US-PP calculations were performed 
by using a cutoff of 25-200 Ry for the KS orbitals and density, 
respectively, and a cubic cell of side 19.05~\AA,
without any symmetry constraints.

STOs results were obtained using the ADF program \cite{ADF1,ADF2};
the frozen cores include 1$s$-2$p$ states for Mn, 1$s$ states for O, N and C.
The KS orbitals were expanded in an uncontracted DZ STO,
standard basis set II \cite{basis} for all atoms
with the exception of the transition metal
for which we used a standard basis set IV \cite{basis},
of TZP quality.
STOs calculations were performed within C$_{2v}$ and C$_S$ symmetry
constraints for species {\bf I} and {\bf II}, respectively. 

\subsection{Results}

In Table 5 we compare the relative energy of the singlet ground states
\cite{Groves1} of the  
two isomeric 2-Pyp and 4-Pyp porphyrins, for both the oxo-aquo and 
oxo-hydroxo species. For the PP calculations we report both
uncorrected and Makov-Payne corrected energy differences.
Table 6 contains the calculated values of the quadrupole moment
used in Eq. (\ref{MP}). 

Results obtained with STOs localized basis sets compute the oxo-aquo
4-Pyp system to 
be ca. 26 kcal/mol more stable than the 2-Pyp one. The Makov-Payne
corrected US-PP results are in excellent agreement with STOs
results, while uncorrected PP results indicate instead
that the 4-Pyp isomer is more stable than the 2-Pyp one by
only 9.5 kcal/mol. 
Moreover, in the case of the oxo-hydroxo species, the uncorrected
results yield an incorrect energy ordering, with the 2-Pyp isomer
computed to be more stable than the 4-Pyp one by 7.0 kcal/mol.
The discrepancy is resolved upon correcting the total energies 
with the Makov-Payne term, resulting again in an excellent agreement
with the STOs energy differences.
\begin{table}[ht]
\caption{Energy differences $\Delta E_{2-4}$ (kcal/mol) 
between the two isomeric porphyrins, for the oxo-aquo (first row) 
and oxo-hydroxo (second row) species,
computed with localized STOs basis and with US PPs,
with Makov-Payne (MP) correction and uncorrected.}
\vspace{1cm}
\begin{tabular}{r|r|c|cc}
        &      &      & \multicolumn{2}{c}{US PPs} \\
       &  & STOs & MP & no-MP \\ \hline 
{\bf I} & $\Delta E_{2-4}$  & 26.4 & 24.5 & 9.5 \\
{\bf II} & $\Delta E_{2-4}$ & 11.5 & 11.1 & -7.0 \\ \hline
\end{tabular} 
\end{table}

\begin{table}[ht]
\caption{Quadrupole moments (a.u.) calculated
with US PPs for the 2-Pyp ($Q_2$) and 4-Pyp ($Q_4$) isomers 
of the oxo-aquo (first row) and oxo-hydroxo (second row) 
species.}
\vspace{1cm}
\begin{tabular}{r|cc}
           &  $Q_2$ & $Q_4$ \\ \hline
{\bf I}  & 518.1 & 627.2  \\
{\bf II} & 412.8 & 522.8  \\
\end{tabular} 
\end{table}
\begin{table}[ht]
\caption{Main optimized geometrical parameters (\AA, degree)
for the 2-Pyp and 4-Pyp oxo-aquo isomers ({\bf I}),
computed with localized STOs basis and with US PPs.
}
\vspace{1cm}
\begin{tabular}{l|c|c|c|c}
 &  \multicolumn{2}{c|}{${\bf 2-Pyp}$ }& \multicolumn{2}{c}{${\bf 4-Pyp}$} \\ \hline
 & US PPs & STOs & US PPs & STOs \\  \hline
r$_{Mn \equiv O}$  & 1.54& 1.57 & 1.54 & 1.57\\
r$_{Mn-OH_2}$  & 2.21 & 2.20& 2.22& 2.21\\
r$_{Mn-N}$$_{av.}$  & 2.04& 2.02 & 2.04& 2.02 \\
$\angle$ N-Mn-N$_{par}$ & 163.3 & 163.5& 163.4& 163.8\\
$\angle$ N-Mn-N$_{per}$ & 167.8& 166.6 & 166.9& 165.4\\
\end{tabular} 
\end{table}

Interestingly, the geometrical structures of the investigated
charged systems calculated using
the US-PP approach with PBC,
turn out to quantitatively compare with results
obtained using localized basis sets, see Table 7 for a comparison of
main optimized geometrical parameters of species {\bf I}. The main
discrepancy (0.03~\AA) is computed for the formally triple Mn$\equiv$O
bond, probably because of the lack of polarization functions in the O
STO basis set which in turn leads to an overestimate of such
parameter. 
The agreement between US PPs and STOs results suggests
that the error introduced
on the electrostatic potential by the presence of a charge in PBC does
not significantly affect the structural properties. 
On the other hand, the effect on the total 
energy is sizable but mostly corrected by the use of Eq. (\ref{MP}).

\subsection{Conclusions}

We believe that our results demonstrate that the Car-Parrinello
approach in conjunction with ultrasoft pseudopotentials represents
a valuable and relatively cheap tool to describe the electronic 
and geometrical properties of complex bio-inorganic
systems, including highly charged and open-shell species.
The study of the electronic and geometrical properties of 
such systems can now be achieved at a reasonable computational cost 
on conventional parallel machines with a limited number 
of processors. Moreover, a simple correction allows us to calculate 
energy differences in charged systems with an accuracy
that is comparable to that of localized basis-set
calculations not using Periodic Boundary Conditions.
First-principle molecular dynamics simulations
of extended bio-inorganic systems may 
still be feasible only on state-of-the art massive parallel computer.
The key to reach such a goal is the 
availability of parallel machines with increased performances 
and number of processors {\em and} highly optimized scalable
algorithms.
We believe that the present parallel implementation of the
Car-Parrinello method using
ultrasoft pseudopotentials
provides such an algorithm
that will allow in the near future the simulation of the 
dynamical properties of such complex systems. 

{\bf Acknowledgments}
Calculations were performed at Keck computational facility of the
Princeton Materials Institute and at ISTM.
Support from the NSF (CHE-0121432) is
acknowledged. We wish to thank Prof. J. T. Groves, Prof. T. G. Spiro 
and Dr. A. Jarzecki for helpful discussions.
FDA thanks CNR (Progetto Finalizzato ``Materiali Speciali
per Tecnologie Avanzate II'') for financial support.
PG thanks MIUR grant PRIN 2001-028432 for partial support.


\end{document}